\documentclass{svproc}
\usepackage{url}
\usepackage{graphicx}

\usepackage{hyperref}

\begin{document}
\mainmatter              
\title{Determination of Chemical Freeze-out Parameters from Net-kaon Fluctuations at RHIC }
\titlerunning{Det. of Chemical FO parameters from net-K fluctuations at RHIC}  
%
\author{Jamie M. Stafford\inst{1} \and Paolo Alba\inst{2} \and Rene Bellwied\inst{1} \and
Valentina Mantovani-Sarti\inst{3} \and Jacquelyn Noronha-Hostler\inst{4} \and Paolo Parotto\inst{1,5} \and Israel Portillo-Vazquez\inst{1} \and
Claudia Ratti\inst{1} }
\authorrunning{Jamie M. Stafford et al.} 
%
\tocauthor{Ivar Ekeland, Roger Temam, Jeffrey Dean, David Grove,
Craig Chambers, Kim B. Bruce, and Elisa Bertino}
\institute{University of Houston, Houston, TX 77204, USA,\\
\email{jmstafford@uh.edu},
\and
Lucht Probst Associates GmbH, Grosse Gallusstrasse 9, D-60311 Frankfurt am Main, Germany
\and
Technische Universit\"at M\"unchen,
James Franck Strasse 1, 85748 Garching, Germany
\and
University of Illinois at Urbana-Champaign, Urbana, IL 61801, USA
\and
University of Wuppertal, Wuppertal D-42097, Germany}

\maketitle              

\begin{abstract}
We calculate the mean-over-variance ratio of the net-kaon fluctuations in the Hadron Resonance Gas (HRG) Model for the five highest energies of the RHIC Beam Energy Scan (BES) for different particle data lists. We compare these results with the latest experimental data from the STAR collaboration in order to extract sets of chemical freeze-out parameters for each list. We focused on the PDG2012 and PDG2016+ particle lists, which differ largely in the number of resonant states. Our analysis determines the effect of the amount of resonances included in the HRG on the freeze-out conditions.
\end{abstract}
\section{Introduction}
Characterizing the transition region of the QCD phase diagram is a focal point for current theoretical and experimental investigations in nuclear physics. Lattice QCD calculations have shown that there is a cross-over transition at T $\simeq$ 155 MeV in the low $\mu_B$ region of the phase diagram \cite{Aoki2006,Borsanyi2010,Bazavov2012}. In addition, models have indicated that the transition becomes first order at high baryonic chemical potential, thus implying the presence of a critical point \cite{Bzdak2019}. The search for this proposed critical point represents the main goal of the Beam Energy Scan II (BES-II) program at RHIC. 
The evolution of the system in a heavy-ion collision (HIC) can be  broken into several stages; two important stages of HICs, the chemical and kinetic freeze-outs, can be related to the experimental results. The chemical freeze-out corresponds to the point in the evolution of the system at which inelastic collisions cease, and at this time the chemical composition is fixed, corresponding to the measured particle yields and fluctuations. The next freeze-out stage, the kinetic freeze-out occurs when the system is so sparse that elastic collisions can no longer occur, and this corresponds to information on the particle spectra and correlations. Within this framework it is clear that the study of the chemical freeze-out is an important aspect of phenomenological explorations in heavy-ion collisions. 
Chemical freeze-out parameters are typically obtained by treatment of the particle yields or fluctuations in a thermal model, such as the Hadron Resonance Gas (HRG) model \cite{Acharya2018,Alba2014,Andronic2009}. Thermal fits of particle yields can be used to determine the temperature, baryonic chemical potential, and volume at freeze-out (${T_f, \mu_{B,f}, V_f}$). In addition, the fluctuations of conserved charges can be used to identify freeze-out parameters by comparing experimental results for the particle fluctuations to a thermal model. The HRG Model has been used to determine the chemical freeze-out conditions in this way in works such as \cite{Alba2014}. This study produced sets of freeze-out parameters for a range of collision energies of the BES by performing a combined fit of net-p and net-Q. Thus, these correspond to the freeze-out conditions of the light particles. Combining this information with the isentropic trajectories from Lattice QCD, we study the net-kaon fluctuations in the HRG model in order to determine if the kaons freeze-out at the same temperature as the light hadrons. This addresses an important effect that has been seen in data at both RHIC and the LHC, which is the tension between the light and strange particles in thermal fits \cite{Acharya2018,Adamczyk2017}. One explanation that attempts to resolve the difference in temperatures is that there could be missing resonances in the thermal model \cite{Noronha-Hostler2014,Bazavov2014}. While recent lattice QCD calculations \cite{Bazavov2014,Alba2017} indicate that several resonances are indeed missing, their full decay channels need to be implemented in order to determine their influence on the freeze-out temperature. Other explanations include a higher freeze-out temperature for strange hadrons \cite{Bellwied2013,Bluhm2019}, pion-nucleon interactions in the S-matrix formalism \cite{Andronic2019}, and large annihilation cross-sections that lead to a lower (anti)proton freeze-out \cite{Steinheimer2013,Becattini2013,Becattini2017} In this study, we seek to characterize the chemical freeze-out during heavy-ion collisions by utilizing the HRG Model to calculate fluctuations of conserved charges.

\section{Methodology}
The Hadron Resonance Gas (HRG) Model describes an interacting gas of ground state hadrons as a system of non-interacting particles and their resonant states. The pressure of such a system of hadrons can be treated in the grand canonical formalism: 
$$\frac{P}{T^4} = \frac{1}{VT^3} \sum_i \, ln \,\! Z_i (T, V, \vec{\mu}),$$

$$\ln \,\! Z^{M/B}_i  = \mp \frac{V \,\! d_i}{(2\pi)^3} \int \!\! d^3 \! k \, \ln \! \left( 1  \mp \, exp \left[ - \left(\epsilon_i - \mu_a X_a^i \right)/T \right] \right)$$
where, the index i runs over all the particles included in the HRG model from the Particle Data Group listing, the energy $\epsilon_i = \sqrt{k^2 + m^2_i}$, conserved charges $\vec{X_i} = \left( B_i, S_i, Q_i \right)$, degeneracy $d_i$, mass $m_i$, and volume $V$. The fluctuations of conserved charges are defined as the derivative of the pressure with respect to the chemical potentials of the conserved charges of interest. 
$$\chi^{BSQ}_{ijk} = \frac{\partial^{i+j+k} \left( P/T^4 \right)}{\partial \left(\mu_B / T \right)^i \, \partial\left(\mu_S / T \right)^j \, \partial \left(\mu_Q / T \right)^k}$$
In order to compare the HRG model results to the experimental data, it is important to also consider the experimental cuts in rapidity and transverse momentum. Taking this into account, the fluctuations of net-kaons, the difference of kaons minus anti-kaons, can be written as:
$$\chi_n^{net-K} = \sum_i^{N_{HRG}} \frac{(Pr_{i \rightarrow net-K})^n}{T^{3-(n-1)}}  \frac{S_i^{1-n}d_i}{4\pi^2} \frac{\partial^{n-1}}{\partial\mu_S^{n-1}} \Bigg\{ \int_{-0.5}^{0.5} d\textit{y} \int_{0.2}^{1.6} d\textit{p}_T \times  $$


$$ \frac{p_T \sqrt {p_T^2 + m_i^2} Cosh[y]}{(-1)^{B_{i}+1} + exp((Cosh[y]\sqrt{p_T^2 + m_i^2} - (B_i\mu_B + S_i\mu_S + Q_i\mu_Q))/T) } \Bigg\} $$ \\
An important goal of this study is to determine the effect of the amount of resonances included in the HRG model on the net-kaon fluctuations. We do so by utilizing different particle lists. In a previous analysis \cite{Alba2017}, it was determined that the most experimentally well-known states are not enough to complete the hadronic spectrum in HRG model calculations as compared to results from Lattice QCD. Therefore, we utilize the measurements of all observed hadronic states from the Particle Data Group over all confidence levels \cite{Patrignani2016}. This compilation of resonances was found to be the one which best reproduces the Lattice QCD results. We will compare the net-kaon fluctuations calculated using this optimal PDG list to the results from \cite{Bellwied2019}, which utilized a different particle list. These lists differ largely in the number of resonant states, particularly in the strange sector.
\section{Results}
\begin{figure}[ht!]
\begin{tabular}{c c}
    \centering
    \includegraphics[width=0.48\textwidth]{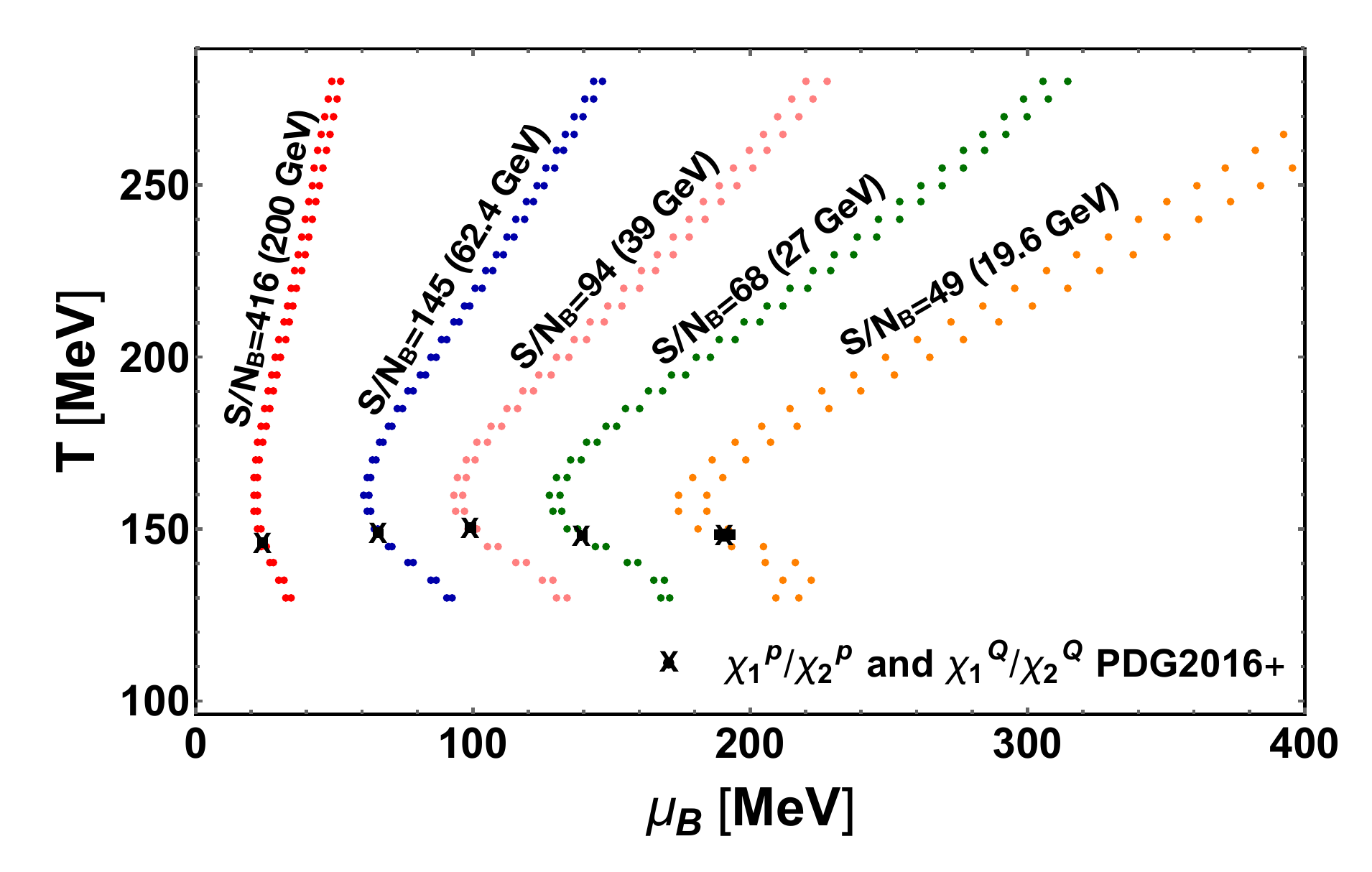} &
    \includegraphics[width=0.48\textwidth]{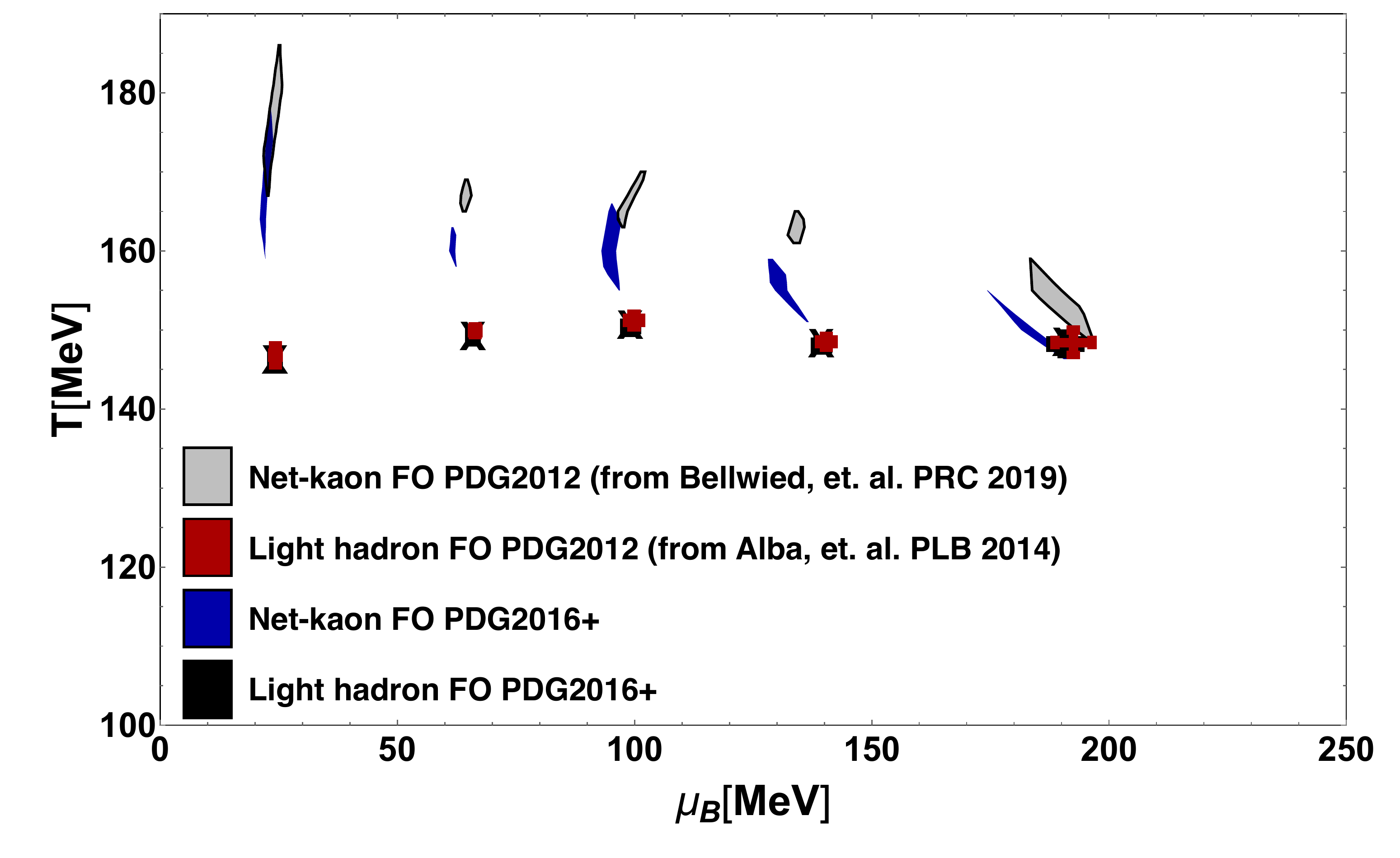}
\end{tabular}
    \caption{(Left) Isentropic trajectories for the five highest energies of the BES. Entropy per baryon number is conserved along these trajectories, which start from the freeze-out parameters determined by the combined fit of $\chi_1^p/\chi_2^p$ and $\chi_1^Q/\chi_2^Q$. (Right) Comparison of net-kaon and light hadron freeze-out parameters for two different HRG particle lists, PDG2012 and PDG2016+, over a range of collision energies at RHIC.}
    \label{fig:isentropesandkaons}
\end{figure}

In order to compare the results from these two different particle lists, we followed the same treatment from \cite{Bellwied2019,Alba2014,Guenther2017}. First, the freeze-out parameters for the light particles were calculated via a combined fit of $\chi_1^p/\chi_2^p$ and $\chi_1^Q/\chi_2^Q$ in the HRG model with the PDG2016+ particle list. Isospin randomization was taken into account for the determination of the freeze-out parameters from net-proton and net-electric charge, as a similar analysis has shown previously \cite{Alba2014}. Then, isentropic trajectories were developed by starting from these new freeze-out points obtained with the PDG2016+ list for the various BES energies and fixing the entropy per baryon number along the path in the phase diagram, consistent with \cite{Guenther2017}. They are shown in figure \ref{fig:isentropesandkaons} (left).

By calculating the net-kaon fluctuations along these isentropes and comparing to the experimental results for the mean-over-variance, we have identified sets of freeze-out parameters, ($T_{FO}$, $\mu_{B,FO}$), for the five highest energies of the Beam Energy Scan. The freeze-out parameters for the kaons with the two different PDG lists are compared to the freeze-out parameters for the light particles determined by the combined fit of net-proton and net-electric charge as shown in figure \ref{fig:isentropesandkaons} (right). Even with the inclusion of more strange resonances, the separation between freeze-out conditions for net-kaons and light particles remains.

\section{Conclusions}
By utilizing different compilations of resonances from the Particle Data Group, we study the effect of the number of particles included in the HRG model on the results for net-kaon fluctuations. We find a slightly lower freeze-out temperature for the kaons when more strange resonances are included in the particle list. However, at  $\sqrt{s_{NN}}$=200 GeV there is a separation of about 10 MeV between these two sets of freeze-out conditions for net-kaons and light hadrons. 

\section{Acknowledgments}
This material is based upon work supported by the National
Science Foundation under Grant No. PHY-1654219 and by the U.S. Department of Energy, Office of Science, Office of Nuclear Physics, within the framework of the Beam Energy Scan Theory (BEST) Topical Collaboration. We also acknowledge the support from the Center of Advanced Computing and Data Systems at the University of Houston. P.P. also acknowledges support from the DFG grant SFB/TR55. J.N.H. acknowledges the support of the Alfred P. Sloan Foundation, support from the US-DOE Nuclear Science Grant No. de-sc0019175. R.B. acknowledges support from the US DOE Nuclear Physics Grant No. DE-FG02-07ER41521.

%
%
%




\end{document}